\documentclass[epj,twocolumns]{svjour_mod}
\RequirePackage{snapshot} 
\usepackage[utf8]{inputenc}
\usepackage{latexsym}
\usepackage{graphics}
\usepackage{amssymb}
\usepackage[colorlinks,citecolor=blue,urlcolor=blue,linkcolor=blue]{hyperref}
\usepackage{hyphenat}
\usepackage{amsmath}
\usepackage{cite}
\usepackage{relsize}
\usepackage{trimclip}
\usepackage{booktabs} 
\usepackage{tabularx}
\usepackage{multirow}
\usepackage{fixmath}
\usepackage{tikz}
\usetikzlibrary{calc}
\usepackage{placeins}
\usepackage{xspace}
\usepackage{slashed}

\usepackage{color}

\newcommand{\LundNet}{\texttt{LundNet}}
\newcommand{\LundNetT}{\texttt{LundNet3}}
\newcommand{\LundNetF}{\texttt{LundNet5}}
\newcommand{\ParticleNet}{\texttt{ParticleNet}}

\definecolor{semiblue}{rgb}{0.3,0.3,0.8}
\newcommand{\logbook}[2]{}

\makeatletter
\g@addto@macro\bfseries{\boldmath}
\makeatother

\begin{document}

\title{Leveraging universality of jet taggers through transfer learning}

\author{Fr\'ed\'eric A. Dreyer\inst{1}, Rados\l{}aw Grabarczyk \inst{1},
  \and   Pier Francesco Monni\inst{2}}

\authorrunning{F. A. Dreyer, R. Grabarczyk, P. Monni}

\institute{
Rudolf Peierls Centre for Theoretical Physics, Clarendon Laboratory, Parks Road, University of
Oxford, OX1
3PU, UK
\and
CERN, Theoretical Physics Department, CH-1211 Geneva 23, Switzerland
}

\date{}

\abstract{
  A significant challenge in the tagging of boosted objects via
  machine-learning technology is the prohibitive computational cost
  associated with training sophisticated models.
  Nevertheless, the universality of QCD suggests that a large amount
  of the information learnt in the training is common to different
  physical signals and experimental setups.
  In this article, we explore the use of transfer learning techniques
  to develop fast and data-efficient jet taggers that leverage such
  universality.
  We consider the graph neural networks \texttt{LundNet} and
  \texttt{ParticleNet}, and introduce two prescriptions to transfer an
  existing tagger into a new signal based either on fine-tuning all
  the weights of a model or alternatively on freezing a fraction of
  them.
  In the case of $W$-boson and top-quark tagging, we find that one can
  obtain reliable taggers using an order of magnitude less data with a
  corresponding speed-up of the training process.
  Moreover, while keeping the size of the training data set fixed, we
  observe a speed-up of the training by up to a factor of three.
  This offers a promising avenue to facilitate the use of such tools
  in collider physics experiments.  }

\PACS{{12.38.-t}{Quantum Chromodynamics, Machine Learning}} 
\maketitle

\section{Introduction}
\noindent
The tagging of energetic heavy particles through machine learning
methods is one of the key technical challenges at the Large Hadron
Collider.
Such identification techniques are used to search for new-physics
signatures (see e.g. Refs.~\cite{Collins:2018epr,Heimel:2018mkt,DeSimone:2018efk,DAgnolo:2018cun,Collins:2019jip,Aguilar-Saavedra:2020uhm,Kasieczka:2021xcg,Karagiorgi:2021ngt}),
or to study the properties of Standard Model
particles~\cite{Radovic:2018dip}, notably to identify boosted
electroweak
bosons~\cite{deOliveira:2015xxd,Barnard:2016qma,Louppe:2017ipp,CMS:2020poo},
the Higgs
boson~\cite{Lin:2018cin,Lim:2018toa,Datta:2019ndh,Moreno:2019neq,Chakraborty:2019imr,Khosa:2021cyk,Cavallini:2021vot,Qu:2022mxj},
or to assign jet flavour~\cite{Almeida:2015jua,Guest:2016iqz,Komiske:2016rsd,ATLAS:2017dfg,Kasieczka:2017nvn,Butter:2017cot,ATLAS:2017gpy,CMS:2017wtu,Cheng:2017rdo,Macaluso:2018tck,Moreno:2019bmu,Andreassen:2020nkr,ATLAS:2020jip,Kasieczka:2020nyd,dAgnolo:2021aun,Romero:2021qlf,Dreyer:2021hhr}.
The most challenging scenario is the one in which such heavy objects
decay into hadronic jets, in which case the ability to identify them
from the decay products is seriously challenged by the overwhelming
background arising from QCD jets.
Provided one has a robust theoretical control over such background
processes, the use of pattern-recognition methods from computer
science can help construct novel taggers with a significantly improved
performance with respect to analytic discriminants (see
e.g. Ref.~\cite{review,reviewexperimental, Larkoski_2020} for recent
reviews).

A large variety of methods has been proposed in recent years,
achieving a remarkable performance in discriminating signal from
background in different experimental measurements.
These are based on several types of techniques, which span from the
use of theory-motivated observables such as energy-flow
polynomials~\cite{Komiske:2017aww}, convolutional neural
networks~\cite{deOliveira:2015xxd} and graph
networks~\cite{Qu:2019gqs} that use four-momenta as input variables.
Among the more recent tools, \LundNet{}~\cite{LundNet} combines the
performance of the state of the art graph networks with
theory-motivated kinematic input variables, namely the Lund jet
plane~\cite{Andersson:1988gp} of emissions within a
jet~\cite{Dreyer:2018nbf}.

In the application of machine learning technology to jet tagging, a
first challenge is represented by the robust assessment of the
theoretical uncertainty in a given model.
This is dominated by the dependence of the model's ability to
discriminate a given signal from the QCD background on the underlying
simulation that is used in the training.
A precise control over these effects demands the development of more
accurate event generators, a task that is receiving significant
attention in the literature (see e.g. Refs.~\cite{Hamilton:2013fea,
  Alioli:2013hqa,Hoeche:2014aia,Monni:2019whf,Monni:2020nks,Alioli:2021qbf,
  Campbell:2021svd,Prestel:2021vww,Hoche:2017hno,Dasgupta:2018nvj,Bewick:2019rbu,
  Dasgupta:2020fwr,Forshaw:2020wrq,Nagy:2020rmk} and references
therein for some recent developments).
A second outstanding challenge is reducing the high computational cost
of training sophisticated models, which besides the generation of
large samples of events also usually requires running on GPUs for
several days. In consequence, the training of such taggers requires
computational resources not always in the reach of their potential
users. Moreover, the models are highly dependent on the experimental
signal (e.g. $W$ vs. top-quark tagging) as well as on the choice of
experimental cuts which makes the training of a tagger for specific
experimental needs from scratch highly inefficient.
The physical picture suggests, however, that most of the information
learnt by a tagger is related to the description of the QCD splittings
that occur within a jet, which simply encode universal properties of
QCD rather than features that depend on the underlying experimental
signal.
It has been demonstrated multiple times in the past that early layers
of a deep convolutional network extract general features from the
data, and can thus be potentially reused for new
tasks~\cite{DBLP:journals/corr/abs-1808-01974}.
In the present short article we examine the latter aspect of jet
tagging, and we tackle the problem by applying inductive \textit{transfer
  learning} techniques~\cite{articletransfer} to leverage an existing
model for a new application to a different experimental signature.
As a result, we will discuss the construction of computationally
efficient jet taggers that can achieve high performance also when
trained on a small fraction of the original data set, with a
significant reduction in the computational complexity associated with
the training.
The article is structured as follows. In Sec.~\ref{sec:lundnet} we
briefly review the graph neural network \LundNet{} that we adopt
for our studies, and discuss the underlying description of jets in
terms of Lund jet plane declusterings.
In Sec.~\ref{sec:transfer} we then introduce two transfer-learning
procedures that allow one to train a new signal starting from an
existing model trained on a different tagging problem.
These techniques are then applied to the problem of top tagging in
Sec.~\ref{sec:case-study-top}, where we study in detail the
performance of transfer learning between top taggers with different
transverse-momentum cuts and from a $W$ tagger to a top tagger. Subsequently, we
present an analysis of the computational advantages of transfer
learning procedures over training new models from scratch.
In Sec.~\ref{sec:conclusions} we discuss our conclusions.

\section{Graph neural networks in the Lund plane}
\label{sec:lundnet}
\begin{figure}[h]
\centering
\includegraphics[width=0.8\linewidth]{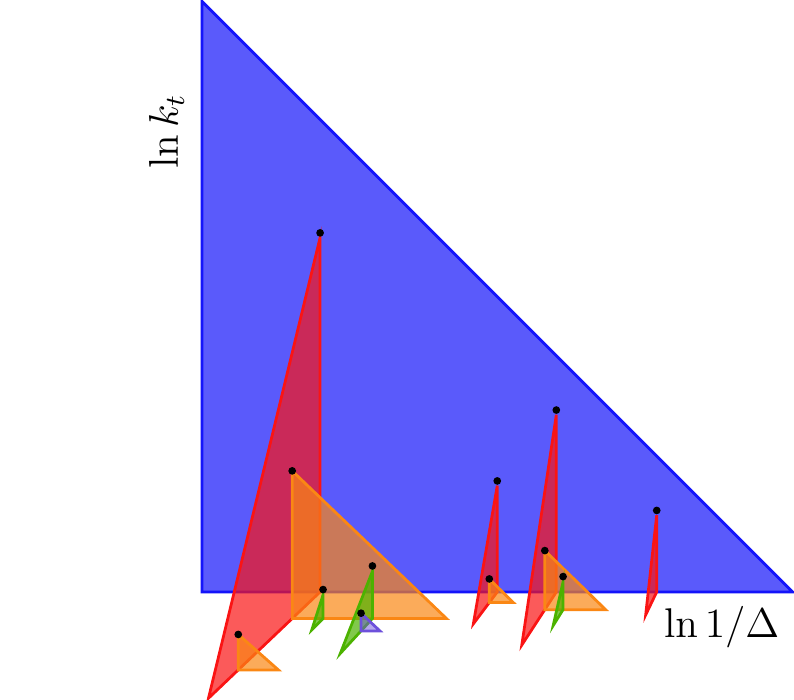}
\caption{A graphic representation of the Lund plane for the radiation
  within a jet. The blue triangle represents the primary Lund plane,
  with secondary and tertiary Lund planes shown in red and orange,
  respectively.}
\label{fig:LJP}
\end{figure}

\noindent
The Lund jet plane~\cite{Dreyer:2018nbf} is a useful theoretical
framework to represent the internal kinematics of a jet by means of
Lund diagrams~\cite{Andersson:1988gp}.
To define it, one starts by constructing the
Cambridge-Aachen~\cite{Dokshitzer:1997in,Wobisch:1998wt} clustering
sequence using the constituents of the jet, which carries out a
sequential pair-wise recombination of the two proto-jets with the
smallest angular separation in rapidity-azimuth.
One maps this clustering sequence to a tree of Lund declusterings,
each of which encodes the kinematic properties of the corresponding
clustering step. Each declustering $p_i \to p_a, p_b$ can be
parametrised in terms of the following set of variables:
\begin{align}
  \label{eq:branching-variables}
  \Delta &\equiv \Delta_{ab},\quad
           k_t \equiv p_{Tb} \Delta_{ab},\quad
           m^2 \equiv  (p_a + p_b)^2,\notag
  \\
  z &\equiv \frac{p_{Tb}}{p_{Ta}\!+\!p_{Tb}},
      \quad
      \psi \equiv \tan^{-1} \frac{y_b\!-\!y_a}{\phi_b \!-\! \phi_a},
\end{align}
where $p_a$, $p_b$ are the post-branching momenta with their
transverse momenta ordered such that $p_{Tb} < p_{Ta}$,
$\Delta_{ab} = \sqrt{(y_a-y_b)^2 + (\phi_a-\phi_b)^2}$ (with $y$ and
$\phi$ denoting the rapidity and azimuth, respectively), $\psi$ is an
azimuthal angle around the subjet axis, and $z$ is the transverse
momentum fraction of the branching.
The construction of the Lund jet plane can be schematically understood
with the help of Fig.~\ref{fig:LJP}. The (primary) Lund plane
associated with the initial proto-jet represents a two-dimensional
parametrisation of the phase space available to further radiation from
it. This is indicated by the large (blue) triangle in the
$\ln k_t-\ln 1/\Delta$ plane in Fig.~\ref{fig:LJP}.
Each subsequent primary emission along the hard branch of the tree is
shown in red, and it forms a new leaf of the Lund plane, from which
secondary emissions will be radiated, indicated by orange leaves.
The procedure iterates through all branches of the clustering history,
leading to a complete representation of the jet's substructure.
In particular, the structure of the primary Lund jet plane can be
computed accurately with perturbative methods~\cite{Lifson:2020gua}
and measured experimentally~\cite{ATLAS:2020bbn}.

\LundNet{}~\cite{dreyer2021jet} is a graph neural network which
takes the Lund jet plane as input to train efficient and robust jet
taggers. 
The resulting taggers outperform tools with low-level inputs
\cite{dreyer2021jet} and are relatively resilient to non-perturbative
and detector effects given an appropriate choice of cuts in the Lund
plane.
The jet is mapped into a graph whose nodes represent the declustering
steps of the Cambridge-Aachen history, parametrised in terms of tuples
$\mathcal{T}^{(i)}$ containing the kinematic variables defined in
Eq.~\eqref{eq:branching-variables}. In particular, one can define two
versions of the \LundNet{} network based on the dimensionality of the
input tuple, defined as follows:
\begin{align}
  \label{eq:nodes3}
    {\rm \LundNetT:} \quad &\mathcal{T}^{(i)} = \{k_t^{(i)},
                                     \Delta^{(i)}, z^{(i)}\}\,,\\
  \label{eq:nodes5}
    {\rm \LundNetF{}:} \quad &\mathcal{T}^{(i)} = \{k_t^{(i)}, \Delta^{(i)}, z^{(i)}, m^{(i)}, \psi^{(i)}\}\,.
\end{align}
The edges of the graph correspond to the structure of the
Cambridge-Aachen tree.

The \LundNetF{} network contains more kinematic information for
each declustering node, and therefore results in a higher tagging
efficiency. Conversely, the \LundNetT{} network has been shown
to be more resilient to non-perturbative and detector
effects~\cite{dreyer2021jet}, while having an efficiency similar to
state-of-the-art taggers.

\begin{figure}[h]
\centering
\includegraphics[width=\linewidth]{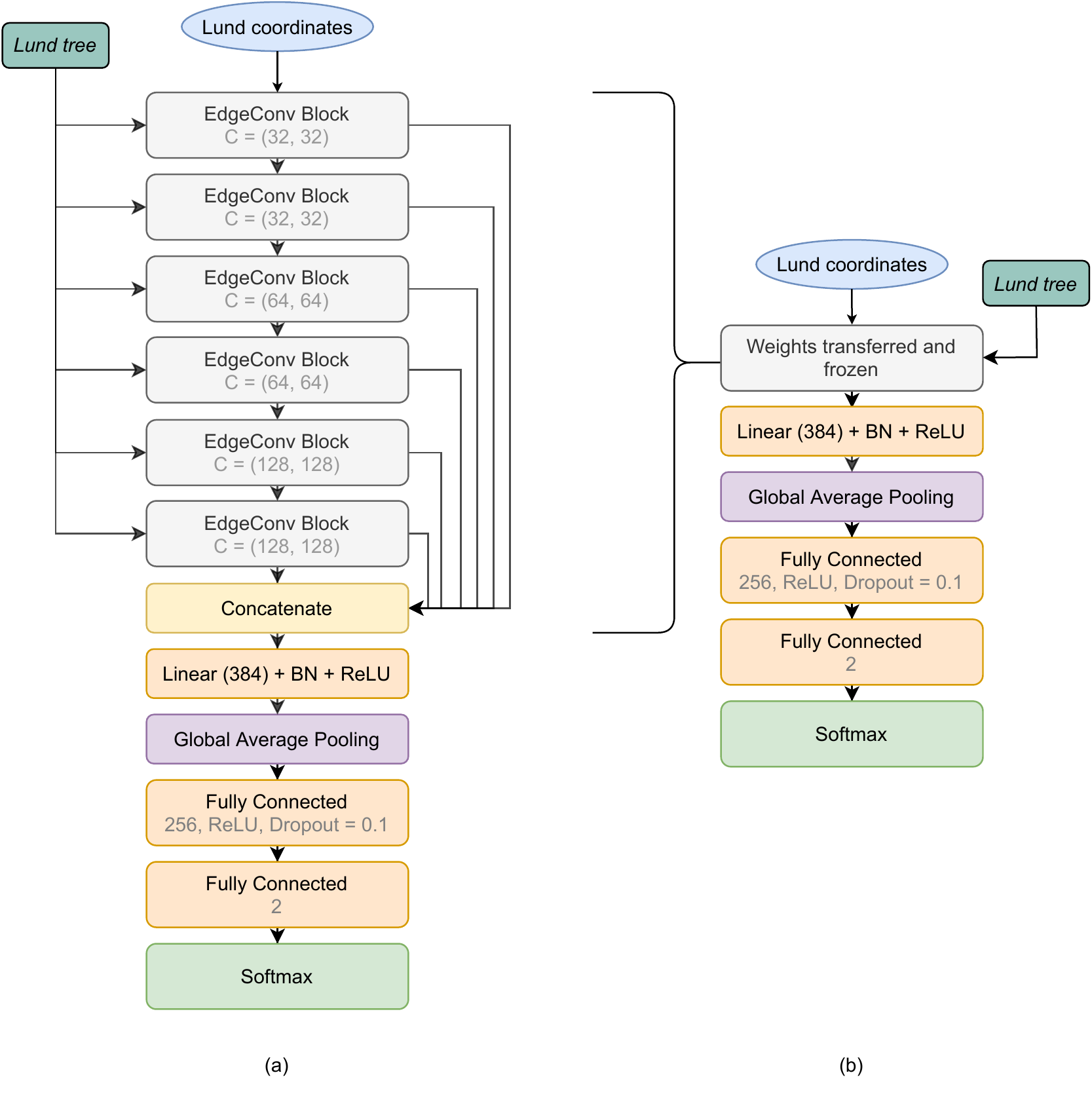}
\caption{A flowchart representing the \LundNet{} architecture
  and the transfer learning procedure employed in this article.}
\label{fig:architectures}
\end{figure}

The core of the graph architecture relies on an EdgeConv
operation~\cite{DGCNN}, which applies a multi-layer perceptron (MLP)
to produce a learned edge feature, using combined features of node
pairs along each edge as input.
This shared MLP consists of two layers, each with a dense network,
batch normalisation~\cite{DBLP:journals/corr/IoffeS15} and ReLU
activation~\cite{glorot2011deep}.
This is followed by an aggregation step which takes an element-wise
average of the learned edge features along the edges.
The model also includes a shortcut connection~\cite{he2016deep}.
The same MLP is applied to each node, updating all node features while
keeping the structure of the graph itself unchanged.
The LundNet architecture consists of six successive EdgeConv blocks,
with the number of channels for each MLP pair in the block given by
$(32, 32)$, $(32, 32)$, $(64, 64)$, $(64, 64)$, $(128, 128)$ and
$(128, 128)$.
Their final output is concatenated, and processed by a MLP
with 384 channels, to which a global average pooling is applied to
extract information from all nodes in the graph.
This is followed by a fully connected layer with 256 units and a
dropout layer with rate 10\%.
A final softmax output provides the result of the classification.
This model is implemented with the Deep Graph Library
0.5.3~\cite{wang2020deep} and the PyTorch~1.7.1
\cite{NEURIPS2019_9015} backend, using an Adam
optimiser~\cite{DBLP:journals/corr/KingmaB14} to minimise the cross
entropy loss.
The LundNet architecture is summarised in
figure~\ref{fig:architectures}a.
Training is performed for 30 epochs using an initial learning rate of
0.001, which is lowered by a factor 10 after the 10th and 20th epochs.

The architecture of \LundNet{} is based on a similar graph neural
network, \ParticleNet{}~\cite{Qu:2019gqs}, which also provides
excellent performance on LHC classification tasks, and we will use it
as one of the benchmarks in our study below in comparison to
\LundNet{}.

\noindent

\section{Transfer learning}
\label{sec:transfer}
In this section we briefly discuss the application of transfer
learning techniques to the design of jet taggers.
Transfer learning aims at reusing pre-trained models on new problems,
leveraging the knowledge obtained on a similar task to improve the
training of a new model, for example by using some or all of the
weights of an existing pre-trained neural network as starting point.

Transfer learning has seen a wide range of applications, notably in
language processing and computer vision~\cite{5288526,DBLP:journals/corr/abs-1911-02685}.
While earlier data-based approaches to transfer learning focused on
domain
adaptation~\cite{DBLP:journals/corr/abs-0907-1815,li-etal-2012-cross},
there has been a surge of interest in recent years in adapting deep
learning models to new tasks~\cite{DBLP:journals/corr/abs-1808-01974}.
The two main avenues to achieve this goal is either through the
retraining of a deep neural network while freezing the weights of its
initial layers~\cite{DBLP:journals/corr/YosinskiCBL14}, or through the
fine-tuning of the
model~\cite{DBLP:journals/corr/GirshickDDM13,10.5555/3044805.3044879,10.1145/2818346.2830593}.
While most existing applications are based on convolutional or
recurrent neural networks, the development of deep learning on graph
structured data has also seen advances in transfer learning applied on
graph neural
networks~\cite{DBLP:journals/corr/abs-1905-12265,10.5555/3298483.3298550}.

In the context of machine learning applications to jet physics, one
could expect that different taggers rely on a certain amount of
information that is common to different tasks. 
Concretely, the properties of QCD that define the radiation pattern
inside a boosted jet stemming from the QCD background is largely
identical within different taggers, and further commonalities can be
identified among signals with a similar number of prongs produced by
the resonance decay inside a (fat) jet.
This suggests that jet physics is an ideal area for the application of
transfer learning methods.
On the practical side, this would allow for the design of new
models/taggers starting from a pre-existing one, not necessarily
trained on the same task. 
The main advantage of transfer learning would then be the considerably
reduced computational cost associated with the training of the new
model, which does not need to be built from scratch for every new
task.

A first important question is the extent to which a network is
\textit{transferable}, i.e. whether the transferred model is capable
of reaching a performance that is as close as possible to the fully
trained model with just a fraction of the computing resources.
A general answer to this question requires a thorough investigation of
the features of a given network that are connected to a higher
\textit{transferability}, and this goes beyond the scope of this
article.
Here we instead take a first step in this direction, and consider the
two graph neural networks which have been discussed in the previous
section.
Of these, \texttt{ParticleNet} relies on the information carried by
the full four momenta of the jet constituents, whereas the
\texttt{LundNet} models essentially map the Cambridge/Aachen sequence
of a jet into its own Lund jet plane.
The latter representation encodes the kinematic information at each of
the branchings in the fragmentation process, which in turn is to a
large extent universal across taggers and depends mainly on the
properties of QCD near the soft and/or collinear limits.
A second interesting property of the Lund jet plane is that QCD
(background) jets roughly have a uniform density of emissions in the
Lund jet plane, and hence this structure can be learnt very
effectively by a neural network such as \texttt{LundNet}.
Both of the above properties are expected to facilitate transfer
learning in that the input variables on which \texttt{LundNet} relies
already highlight universal properties of QCD jets and allow the
network to distinguish them from those of typical signal jets stemming
from the decay of a boosted heavy object.
For this reason, our expectation is that transfer learning reaches a
rather good performance in the context of this class of models.
On the other hand, in the case of \texttt{ParticleNet}, the model
needs to learn the non-trivial mapping between the information carried
by the final-state four momenta used as input, to the physical
fragmentation process of the jet. This ends up adding an additional
layer of complexity in the training of the network, which is expected
to be reflected in a lower performance of the models trained via
transfer learning.

In order to explore the \textit{transferability} of the models
adopted in this article, we consider two different approaches to
inductive transfer learning.
Our first transfer-learning approach is a frozen-layer model.
Here the weights of the EdgeConv layers of \texttt{LundNet} or
\texttt{ParticleNet} have been pre-trained on a separate jet sample
and are kept fixed during the retraining process.
The final MLP layers are instead reinitialised to random weights and
retrained on a new sample to specialise the tagger to this new pattern
recognition task.
This procedure is shown in figure~\ref{fig:architectures}b, and the
training on a new data set is performed with the same learning rate and
scheduler as the training for the original model.
The second approach is a fine-tuning of all the weights in the original
tagger.
In this case, the learning rate is reduced by a factor ten (or a
factor three when transferring from a $W$ to a top tagger), and the
tagger is retrained with the same number of epochs and scheduler on a
new data set.

The difference in performance between the transferred models and those
trained from scratch probes the ability of a network to learn features
that are common to different tasks and therefore its suitability to
the application of transfer learning techniques.
Moreover, the difference between the frozen-layer and fine-tuning
procedures probes how much of the high-level information learnt by a
network is extracted from the initial layers, which provides insights
on the extrapolating ability of each model.

\section{Case study with top tagging}
\label{sec:case-study-top}
As a case study, we consider the application to top-quark tagging at
the LHC.
We are interested in discriminating top-quark jets with $p_T > 500$
GeV against the QCD background. We apply and analyse the properties of
transfer learning for four different models, \LundNetT{} and its
transfer from a top-tagger with $p_T > 2$ TeV, \LundNetF{} and its
transfer from either a top-tagger with $p_T > 2$ TeV or a $W$-boson
tagger with $p_T > 500$ GeV, and finally \ParticleNet{} and its
transfer from a top-tagger with $p_T > 2$ TeV.
All models presented in this section are trained with events generated
using Pythia 8.223~\cite{pythia}, considering jets defined according
to the anti-$k_t$ algorithm~\cite{Cacciari_2008} with a jet radius
$R=0.8$ and rapidity $|y| < 2.5$.
Signal events are obtained from the simulation of either $WW$ or
$t\bar{t}$ production, with $W$ bosons decaying hadronically, while
the background sample is obtained from QCD dijet events.
The signal and background training data sets consist of $5\times 10^5$
events each. Validation and testing is done on data sets of
$5\times 10^4$ events for each of the signal and background
process.\footnote{The data is available at
  \url{https://github.com/JetsGame/data}}

We start by discussing the reduction in computational complexity that
can be achieved through the use of transfer learning techniques, and
then provide a phenomenological study of top tagging performance for
each model.

\subsection{Computational complexity of transferred models}
\label{sec:performance}

\begin{figure}[h!]
\centering
\includegraphics[width=\linewidth]{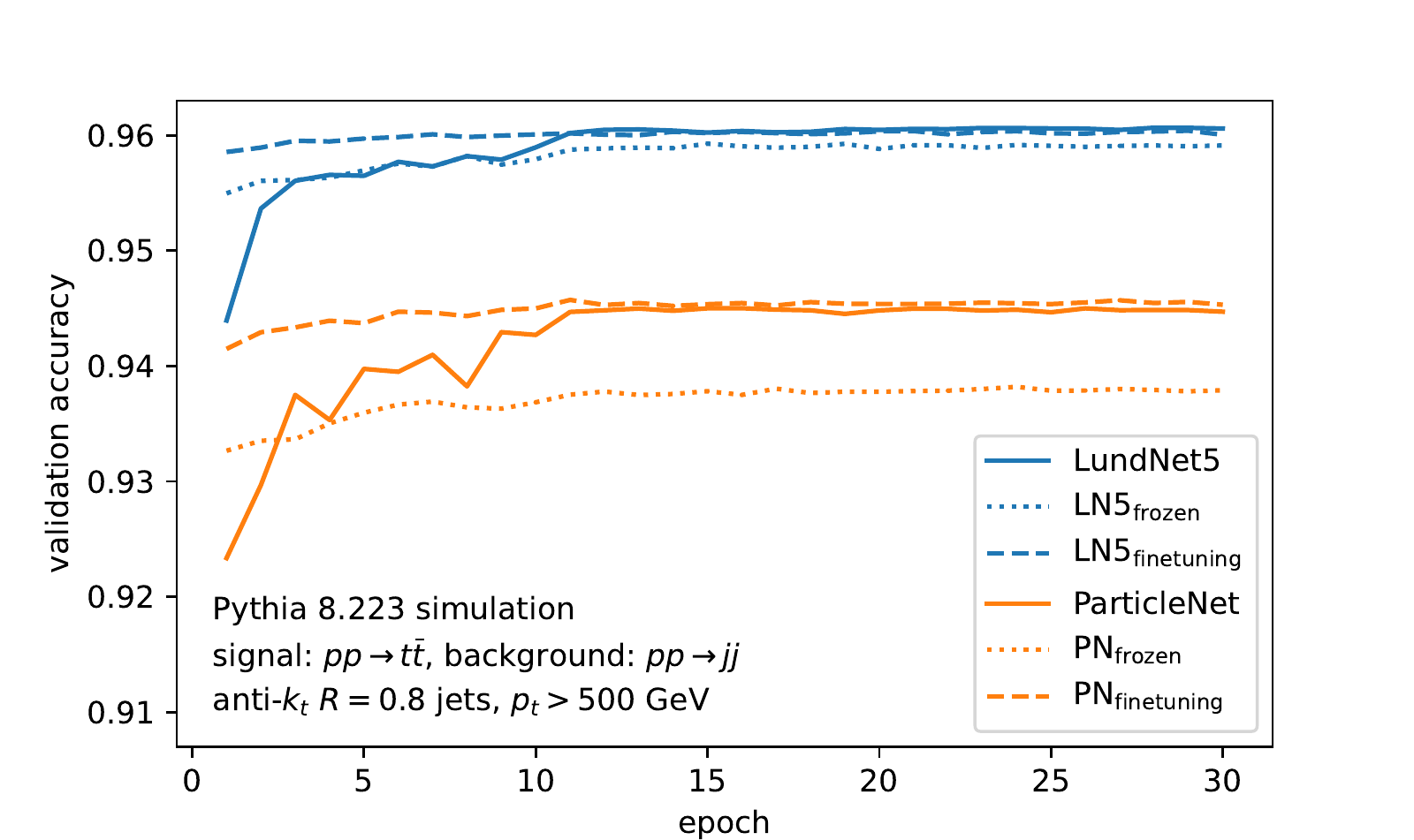}
\caption{Validation accuracy as a function of the number of training
  epochs.}
\label{fig:valacc_epochs}
\end{figure}

As described in section~\ref{sec:transfer}, we consider two transfer
learning approaches to retrain an existing jet tagger, a frozen layer
model and a fine-tuning model.
In this section, we aim to investigate the computational cost of both
approaches.
For this we consider the construction of a top tagger for $500$ GeV
jets, starting from a model trained on $2$ TeV data.
The training time was measured on a NVIDIA GeForce RTX 2080 Ti GPU,
training a tagger on either $10^6$ or $10^5$ total top and QCD jets
with an equal number of signal and background events.
We measure the time required to train a \LundNetF{} and \ParticleNet{}
model,%
\footnote{The training time for both \LundNetF{} and \LundNetT{} is
  almost identical.}  given in milliseconds per sample and epoch (which
is identical for both the $10^6$ and $10^5$ samples) as well as the
corresponding total training time for both data sizes.
These measurements are summarised in Tab.~\ref{tab:timings}.
\begin{table*}[t]
\centering
\phantom{x}\medskip
\caption{Training time for different taggers. The time was measured
  when running the models on an NVIDIA GeForce RTX 2080 Ti card.
  Note that \LundNetF{} transferred from a top tagger with $p_T > 2$
  TeV, when trained on a data set 10 times smaller, still performs
  better than \ParticleNet{} in terms of AUC (see Tab.~\ref{tab:top500GeV}) despite the time of
  training being significantly reduced.  }
\begin{tabular}{lcccccc}
  \toprule
  && Training time&& Total for $10^6$ samples && Total for $10^5$ samples \\
  && [ms/sample/epoch] && [hh:mm:ss] && [hh:mm:ss] \\
 \midrule
 LundNet5 && 0.46 && 03:48:15 && 00:22:43 \\
 LN5$_\text{frozen}$ && 0.15 && 01:17:02 && 00:07:36\\
 LN5$_\text{finetuning}$ && 0.46 && 03:48:32 && 00:22:45\\
 ParticleNet && 3.60 && 30:09:44 && 02:59:17\\
 PN$_\text{frozen}$ && 2.16 && 18:13:21 && 01:47:37\\
 PN$_\text{finetuning}$ && 3.60 && 29:59:46 && 03:01:04\\
 \bottomrule

\end{tabular}
\label{tab:timings}
\end{table*}
All models were trained for 30 epochs, regardless of their
convergence, shown in Fig.~\ref{fig:valacc_epochs} for the full data
sample.
The figure shows the convergence for the two sample cases of
\LundNetF{} (in blue) and \ParticleNet{} (in orange).
The solid lines show the evolution for models trained from scratch,
the dashed lines refer to the fine-tuning transfer learning setup
(trained using a ten-times smaller initial learning rate of
$10^{-4}$), and finally the dotted lines refer to the frozen-layer
transfer learning setup.
%
%
One observes here that, in practice, the transferred models approach
an optimum at a much faster rate, with the fine-tuning setup
converging sooner to higher values of the validation accuracy. This
implies that such models could be trained for only a few epochs to
further reduce the computational cost.

The fine-tuning model does not provide any noticeable speed-up in
training time, as it requires the update of all the weights in the network.
However, as we will see in section~\ref{sec:top-perf}, it can achieve
comparable performance to a full model with only a small fraction of
the data.
As such, if one uses a tenth of the full data sample, this effectively
provides a factor of ten speed-up in training time.
Moreover, as already pointed out, the convergence of the model is
significantly faster, and requires only a few epochs to converge to an
optimal solution, hence providing an opportunity for further
optimisation of the training time.

Conversely, the frozen-layer approach has the advantage of reducing
the computational cost of the retraining by limiting the update of the
weights to the final dense layers, while keeping the EdgeConv blocks
unchanged.
This results in a reduction of the training time by about a factor of
three compared to that of a full model on the same data sample.
As for the fine-tuning model, a further reduction can be achieved by
reducing the number of epochs and the size of the data set.
However, as will be discussed in section~\ref{sec:top-perf}, the
frozen layer model requires a larger data sample than the fine-tuning
approach to achieve comparable performance.

\subsection{Performance of top taggers}
\label{sec:top-perf}

We now study the performance of our top taggers, using the area under
the ROC curve (AUC) as an indicator of a model's performance, and
summarise our results in Tab.~\ref{tab:top500GeV}.
\begin{table*}[t!]
\centering
\caption{Benchmarks for top tagging with $p_T > 500 \textrm{
    GeV}$. The different columns show the AUC for the different
  transfer learning models considered in the text, where FT denotes
  the \textit{fine-tuning} option, FR denotes the
  \textit{frozen-layer} option, and the $(10\%)$ superscript refers to
  results obtained with just one tenth of the original training data.}
\begin{tabular}{ l c c c c c c }
 \hline
  & AUC & AUC$_{\rm FT}$ & AUC$_{\rm FR}$ & AUC$^{(10\%)}$ & AUC$_{\rm FT}^{(10\%)}$& AUC$_{\rm FR}^{(10\%)}$\\
 \hline
 \LundNetT{} (from top 2 TeV)        &0.9820&0.9820&0.9816&0.9773&0.9802&0.9791\\
 \LundNetF{} (from top 2 TeV)        &0.9866&0.9865&0.9863 & 0.9826& 0.9850&0.9845\\
 \LundNetF{} (from $W$ 500 GeV)  &-&0.9863&0.9858& - &0.9834&0.9832\\
 \ParticleNet{} (from top 2 TeV)      &0.9826&0.9826&0.9793&0.9765 & 0.9795&0.9772\\
 \hline
\end{tabular}
\label{tab:top500GeV}
\end{table*}
For each model, we consider the AUCs corresponding to a training from
scratch (indicated by AUC in the table), transfer learning with a
fine-tuning setup (indicated by AUC$_{\rm FT}$), and transfer learning
with a frozen-layer setup (indicated by AUC$_{\rm FR}$).
For the above three options, we also consider the values of the AUC
obtained with $10\%$ of the original training data (denoted by
AUC$^{(10\%)}$ in Tab.~\ref{tab:top500GeV}).
We find that, in the case of \LundNet{} models trained in the full
data set, the fine-tuning setup reproduces exactly the AUC of the model
trained from scratch. We also observe that the frozen option, despite
being considerably cheaper from a computational viewpoint, leads to
AUC values which are extremely close to those of the above models,
indicating that \LundNet{} models are very suitable for the
application of transfer learning as discussed in the previous section.
We also consider the more demanding transfer of a \LundNetF{} $W$
tagger to a top tagger at the same 500 GeV $p_T$ threshold.
We can see from the AUC values shown in Tab.~\ref{tab:top500GeV} that
while a moderate loss of performance is found for the model trained on
the reduced data set, we still recover AUC values for the transferred
top tagger that are very close to the fully trained \LundNetF{} model,
and significantly better than most state-of-the-art taggers.

\begin{figure}[h!]
\centering
\includegraphics[width=\linewidth]{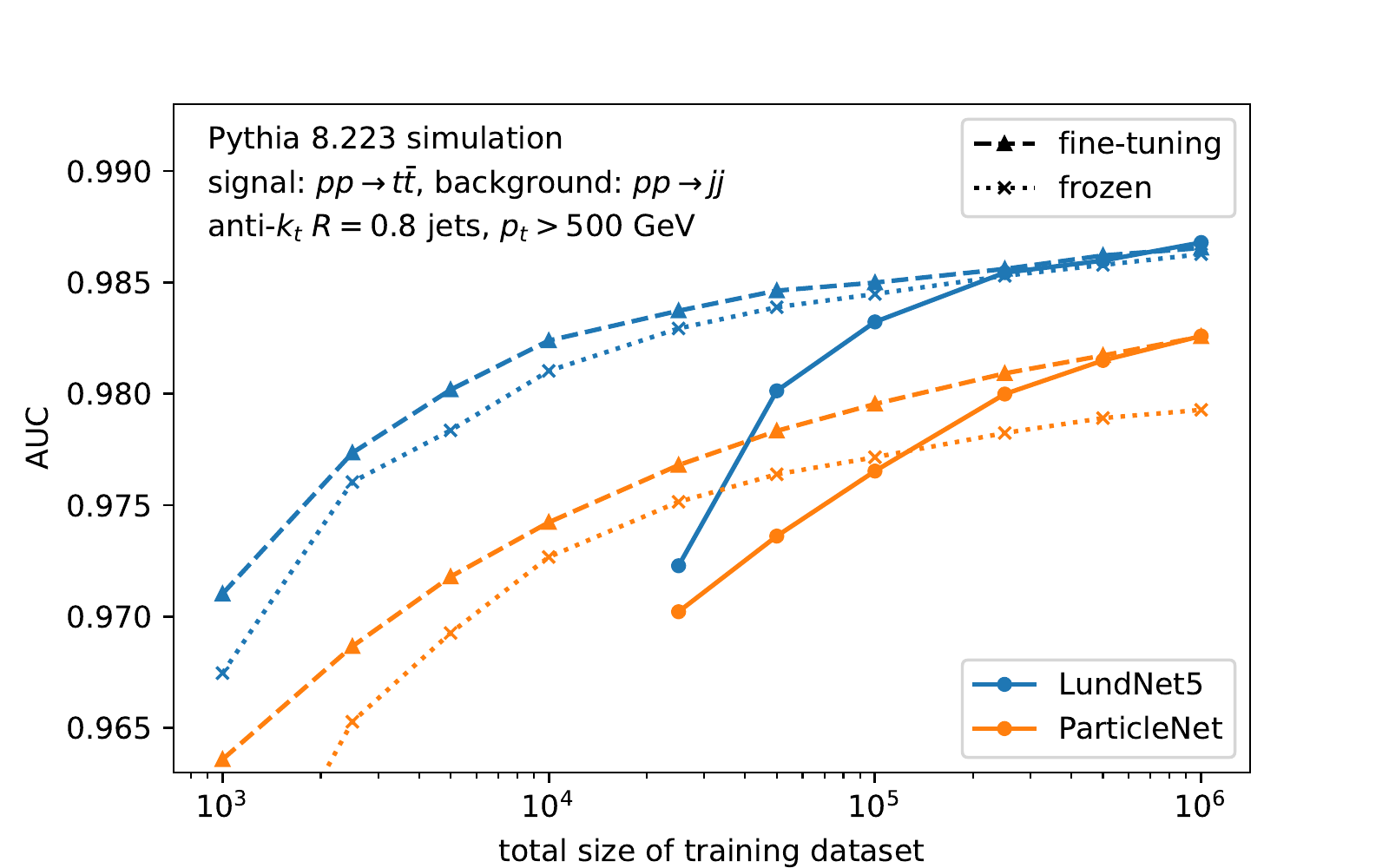}
\caption{Area under the ROC curve as a function of the total signal
  and background training data set size.}
\label{fig:datasetaucs}
\end{figure}

In the case of \ParticleNet{}, the fine-tuning setup still performs as
well as the model trained from scratch while the frozen setup leads to
visibly smaller AUC values.
As expected, this indicates that \ParticleNet{} is less suitable for
the application of transfer learning. This is due to the fact that it
relies on low-level information, such as four momenta, which makes it
less easy to identify general properties of the kinematic pattern of
QCD already in early layers of the network.
For models trained on $10\%$ of the original training data set, we
observe that in the case of \LundNetT{} and \LundNetF{},
the values of AUC obtained with transfer learning models are hardly
affected by the reduction in sample size, and they still perform
nearly as well as the original models trained on the full data set. 
For \ParticleNet{} the performance of the models obtained through
transfer learning is instead closer to that of the model trained (from
scratch) on the reduced data set, in line with our expectation that
this class of models is less transferable.
We also show the dependence of the AUC on the total (signal plus
background) size of the training data set in
Fig.~\ref{fig:datasetaucs}.
The figure shows that transfer learning gives a significant advantage
for small sizes of the training data set.
For example, the retrained \LundNetF{} model with the fine-tuning
setup and $1.25\times 10^4$ events for signal and background data sets
achieves AUC $= 0.983$, meaning that state-of-the-art performance can
be achieved using far smaller data sets than those needed to train a
network from scratch, with a considerable speed-up of the
process. Concretely, when retraining with $2.5\times 10^4$ samples,
the training time is almost two orders of magnitude smaller than that
needed to train a similarly performing \LundNet{} model from scratch.
Importantly, the difference between the fully trained model and the
fine-tuning and frozen-layer transfer learning setups is rather
moderate in the case of \LundNetF{}, which indicates that such class
of models have rather high \textit{transferability} and they can
easily be retrained on a different task.
In the case of \ParticleNet{}, we observe that the fine-tuning setup
still produces AUC values higher than those of the model fully trained
on smaller data sets, although it does not reach the tagging accuracy
observed for \LundNetF{}.
Moreover, Fig.~\ref{fig:datasetaucs} also shows that the performance
of \ParticleNet{} gets significantly worse when using the frozen-layer
setup, with the fully trained model outperforming the transfer
learning results already for a training done on $10^5$ events, while
\LundNetF{} reaches almost the asymptotic values of AUC for this data
sample (see also Tab.~\ref{tab:top500GeV}).
Overall, this clearly shows that the use of transfer learning provides
a promising avenue to reduce the amount of data required to train new
taggers, with certain classes of models such as \LundNet{} being
more suitable for the application of these techniques.
Whether it is possible to define a metric quantifying a priori the
ability of a model to be transferred to a different task with reduced
computational resources than those needed for a full training, and how
to construct better taggers with such features remain interesting open
questions.

\begin{figure}[t!]
\centering
\includegraphics[width=\linewidth]{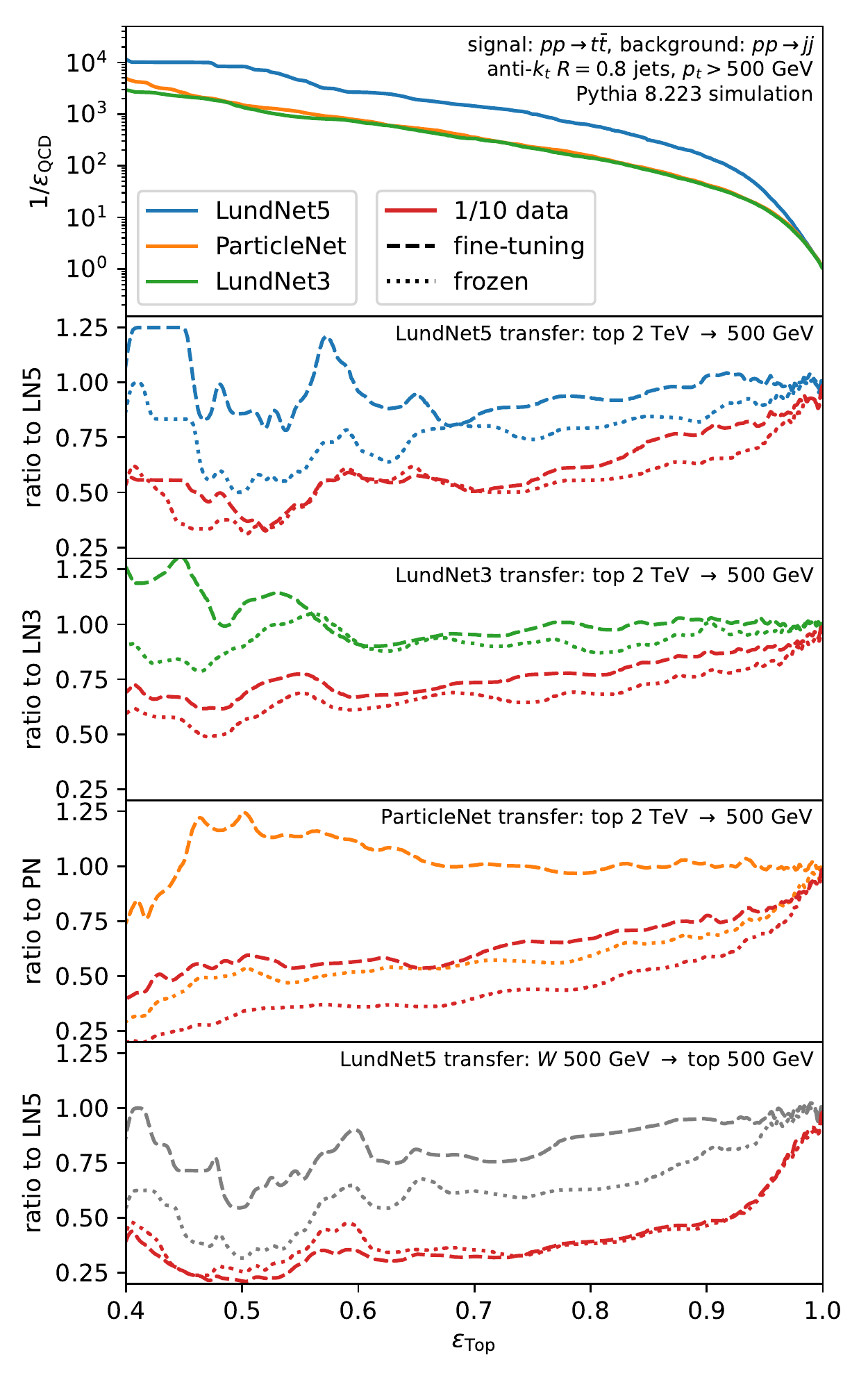}
\caption{QCD rejection vs. Top tagging efficiency.}
\label{fig:Topfrom2TeV}
\end{figure}

We now move on to study the ROC curves corresponding to the different
models in Fig.~\ref{fig:Topfrom2TeV}, showing the background rejection
$1/\varepsilon_{\textrm{QCD}}$ versus signal efficiency,
$\varepsilon_{\textrm{Top}}$. A better performing tagger has a
corresponding ROC curve closer to the top-right corner of the figure.
The upper panel shows the ROC corresponding to the models
\LundNetT{}, \LundNetF{} and \ParticleNet{} all
trained from scratch for a top tagger with $p_T > 500$ GeV. We observe
that, as expected, \LundNetF{} performs better than the other two
models, which achieve a very similar performance. This is due to
the additional information stored in the tuples associated with each
node of the graph (see Eq.~\eqref{eq:nodes5}).
The second panel of Fig.~\ref{fig:Topfrom2TeV} shows the ROC obtained
with \LundNetF{} and different transfer learning options from a
top tagger with $p_T > 2$ TeV, divided by the ROC of the model trained
from scratch (shown in the upper panel).
The dashed blue line corresponds to the fine-tuning setup in which all
weights are re-trained on the new task. This option clearly reproduces
the performance of the tagger trained from scratch, but as already
observed before it does not lead to any reduction of the computational
complexity associated with the training.
The dotted blue line, instead, corresponds to the transfer learning
obtained with the frozen-layer setup which, as already observed in
Tab.~\ref{tab:top500GeV}, leads to a performance that is very close to
that of the original model, with an AUC less than a permille below the
full model, and background rejection at intermediate signal
efficiencies within 20\% of the fully trained tagger.
This performance remains far better than most state-of-the-art jet
taggers, and orders of magnitude above analytic substructure
discriminants.

The remaining three panels in Fig.~\ref{fig:Topfrom2TeV} show a
similar comparison in the case of \LundNetT{} and \ParticleNet{}
models transferred from a top tagger with $p_T > 2$ TeV, and
\LundNetF{} models transferred from a $W$-boson tagger with
$p_T > 500$ GeV.
For the fine-tuned $W$, the initial learning rate is set to
$3\cdot10^{-4}$ to allow for a larger perturbation of the pre-trained
top model.
All of the above four panels also report, in red, the result obtained
with a reduced training data set of $10\%$ of the original size,
i.e. $10^5$ events, with either the fine-tuning (dashed) or
frozen-layer (dotted) setup.
For \LundNet{}, the plot confirms the conclusions drawn from the AUC
study above, showing that these models (both for \LundNetT{} and
\LundNetF{}) still reach the performance of state-of-the-art taggers
also in the transfer learning setups, with the frozen-layer setup
being only moderately less accurate than the computationally more
demanding fine-tuning.
While it is clearly easier to transfer a model from a similar tagger
trained on a different kinematic regime, we see that transfer learning
still reaches highly competitive ROC curves also when the starting
model is a $W$ tagger, shown in the last panel of
Fig.~\ref{fig:Topfrom2TeV}, which demonstrates that the techniques
studied in this article can be adopted across wide families of jet
taggers.
As already observed, \ParticleNet{}, shown in the fourth panel of
Fig.~\ref{fig:Topfrom2TeV}, performs less well under the transfer
learning setups, with a wider gap between the fine-tuning and
frozen-layer options.

In general, from Fig.~\ref{fig:Topfrom2TeV} we conclude that the
retrained models achieve performances which are extremely close to the
models trained from scratch, meaning that the output of EdgeConv
operations is a representation of the data which can be efficiently
reused for other tasks.
The benefits of transfer learning then consist of a significantly
shorter training time and a smaller data set size required to converge
on an efficient tagger. 
All models in the above comparison have been trained for 30 epochs. We
stress once again that the computational cost can be reduced further
by exploiting the fact that transferred \LundNet{} models converge to
an optimum with less epochs, as discussed above in
Fig.~\ref{fig:valacc_epochs}.

\section{Conclusions}
\label{sec:conclusions}
In this article, we have explored the use of transfer learning
methods to train efficient jet taggers from existing models.
With this, we aimed to investigate the ability of a neural network to
learn universal features of QCD and to transfer them to a separate
task.
In practice, we have considered the application of transfer learning to top
tagging at different transverse momentum thresholds and to the tagging
of two- and three-pronged boosted objects, e.g. $W$ boson and top
quark decays.
We studied two jet taggers constructed from graph neural networks,
\LundNet{} and \ParticleNet{}, and conducted a detailed study of the
performance of transferred models as well as of the reduction in
computational complexity provided by transfer learning.

We have implemented two transfer-learning procedures.
The first one relies on fine-tuning all weights in a model by
retraining it on a new data sample with a lower learning rate, while
the second freezes the edge convolutions and retrains solely the final
dense layers of the network.
In the case of \LundNet{} taggers, we find that the fine-tuning
approach requires a similar training time per epoch and sample as the
fully trained model, but converges to an almost optimal solution after
just a few epochs (compared to tens of epochs for a full model) and
requires only a small fraction of the data.
Concretely, a model can achieve nearly the same performance using a
third of the epochs and a tenth of the original training sample, which
leads to a dramatic speed-up of the training process.
On the other hand, the frozen-layer method provides a further speed-up
in training time by a factor three as only a small fraction of the
model weights are updated, but requires a comparatively larger sample
size to achieve a similar performance to the fine-tuning approach.

For the two specific \LundNet{} taggers considered
(\LundNetT{} and \LundNetF{}, which differ in the dimensionality of
the kinematic inputs associated with each node of the graph), we
observe that fine-tuning with a tenth of the data achieves a
background rejection moderately lower than that of a fully trained
model, with the transferred \LundNetT{} tagger recovering slightly
more of the performance of the baseline model.
The frozen-layer approach performs comparably, although in both
\LundNetT{} and \LundNetF{} it achieves slightly lower background
rejection for the same training sample than the fine-tuning method.
%

The conclusions are somewhat different for \ParticleNet{}, where the
frozen-layer method performs noticeably worse than the fine-tuning
approach, regardless of the amount of data and number of epochs.
Furthermore, the background rejection that can be achieved with a
reduced data set is significantly smaller.
We attribute this to the fact that \ParticleNet{} relies on kinematic
information structured as the four momenta of the jet constituents,
which in turn makes it more challenging for the EdgeConv layers to
extract general features about the jet fragmentation.
In comparison, \LundNet{} uses kinematic information of the sequential
clustering steps of the Cambridge/Aachen algorithm as input, which
carries denser information about the jet fragmentation dynamics.
This is reflected in a larger gap between the fine-tuning and
frozen-layer approaches in the \ParticleNet{} case.\sloppy

Our results show that transfer learning constitutes a promising avenue
to build computationally efficient and versatile taggers with
state-of-the-art performance.
This opens a wide array of possibilities for more wide-spread adoption
of machine learning jet-tagging technology for experimental studies at
colliders, such as the Large Hadron Collider and future facilities.
This article provides a first step towards this goal, and motivates
further investigations on the application of these methods to particle
phenomenology.
In this context, a number of interesting theoretical questions arise.
As future directions, it would be informative to study concrete
metrics of transferability of a network, and which features of the
input variables and choices in the architecture of a model can lead to
more transferable designs.
Furthermore, it would be interesting to study knowledge transfer in
jet taggers from first principles, and gain analytical insights into
the behaviour of transferred
models~\cite{Kasieczka_2020,Dreyer:2021hhr}.

\section*{Acknowledgments}
We are grateful to Alexander Huss for constructive comments and
collaboration in the early stages of this work, and to Stefano
Carrazza and Huilin Qu for valuable comments on the manuscript.
This work was supported by a Royal Society University Research
Fellowship (URF$\backslash$R1$\backslash$211294) (FD), and by the
CERN’s Summer Student Programme (RG).

\bibliographystyle{utphys} 
\bibliography{bibfile}

\end{document}